\documentclass[12pt]{article}
\usepackage{latexsym}

\newcommand{\sect}[1]{\setcounter{equation}{0}\section{#1}}

\begin{document}

\title{(Anti-)Evaporation of \\
       Schwarzschild-de~Sitter Black Holes}
\author{{\sc Raphael Bousso}\thanks{\it New address: Department of
       Physics, Stanford University, Stanford, CA 94305-4060;
       bousso1@stanford.edu} \ 
  and {\sc Stephen W. Hawking}\thanks{\it s.w.hawking@damtp.cam.ac.uk}
      \\[1 ex] {\it Department of Applied Mathematics and}
      \\ {\it Theoretical Physics}
      \\ {\it University of Cambridge}
      \\ {\it Silver Street, Cambridge CB3 9EW}
       }
\date{DAMTP/R-97/26\ \ \ \ hep-th/9709224 \\[1ex]
 submitted to Phys.\ Rev.\ D}

\maketitle

\begin{abstract}
  
  We study the quantum evolution of black holes immersed in a
  de~Sitter background space. For black holes whose size is comparable
  to that of the cosmological horizon, this process differs
  significantly from the evaporation of asymptotically flat black
  holes. Our model includes the one-loop effective action in the
  s-wave and large N approximation.  Black holes of the maximal mass
  are in equilibrium. Unexpectedly, we find that nearly maximal
  quantum Schwarzschild-de~Sitter black holes anti-evaporate. However,
  there is a different perturbative mode that leads to evaporation.
  We show that this mode will always be excited when a pair of
  cosmological holes nucleates.

\end{abstract}

\pagebreak

\sect{Introduction}

Of the effects expected of a quantum theory of gravity, black hole
radiance~\cite{Haw74} plays a particularly significant role. So far,
however, mostly asymptotically flat black holes have been considered.
In this work, we investigate a qualitatively different black hole
spacetime, in which the black hole is in a radiative equilibrium with
a cosmological horizon.

The evaporation of black holes has been studied using two-dimensional
toy models, in which one-loop quantum effects were
included~\cite{CGHS,RST1,RST2}.  We have recently shown how to
implement quantum effects in a more realistic class of two-dimensional
models, which includes the important case of dimensionally reduced
general relativity~\cite{BouHaw97a}.  The result we obtained for the
trace anomaly of a dilaton-coupled scalar field will be used here to
study the evaporation of cosmological black holes.

We shall consider the Schwarzschild-de~Sitter family of black holes.
The size of these black holes varies between zero and the size of the
cosmological horizon. If the black hole is much smaller than the
cosmological horizon, the effect of the radiation coming from the
cosmological horizon is negligible, and one would expect the
evaporation to be similar to that of Schwarzschild black holes.
Therefore we shall not be interested in this case. Instead, we wish to
investigate the quantum evolution of nearly degenerate
Schwarzschild-de~Sitter black holes. The degenerate solution, in which
the black hole has the maximum size, is called the Nariai
solution~\cite{Nar51}. In this solution the two horizons have the same
size, and the same temperature. Therefore they will be in thermal
equilibrium. Intuitively, one would expect any slight perturbation of
the geometry to cause the black hole to become hotter than the
background.  Thus, one may suspect the thermal equilibrium of the
Nariai solution to be unstable. The initial stages of such a run-away
would be an interesting and novel quantum gravitational effect quite
different from the evaporation of an asymptotically flat black hole.
In this paper we will investigate whether, and how, an instability
develops in a two-dimensional model derived from four-dimensional
general relativity. We include quantum effects at the one-loop level.

The paper is structured as follows: In Sec.~\ref{sec-cosmo} we review
the Schwarzschild-de~Sitter solutions and the Nariai limit. We discuss
the qualitative expectations for the evaporation of degenerate black
holes, which motivate our one-loop study. The two-dimensional model
corresponding to this physical situation is presented in
Sec.~\ref{sec-2d-model}, and the equations of motion are derived. In
Sec.~\ref{sec-stability} the stability of the quantum Nariai solution
under different types of perturbations is investigated. We find, quite
unexpectedly, that the Schwarzschild-de~Sitter solution is stable, but
we also identify an unstable mode. Finally, the no-boundary condition
is applied in Sec.~\ref{sec-nbp} to study the stability of
spontaneously nucleated cosmological black holes.

\sect{Cosmological Black Holes} \label{sec-cosmo}

\subsection{Metric}

The neutral, static, spherically symmetric solutions of the Einstein
equation with a cosmological constant $\Lambda$ are given by the
Schwarzschild-de~Sitter metric
\begin{equation}
ds^2 = - V(r) dt^2 + V(r)^{-1} dr^2 + r^2 d\Omega^2,
\end{equation}
where
\begin{equation}
V(r) = 1 - \frac{2\mu}{r} - \frac{\Lambda}{3} r^2;
\end{equation}
$d\Omega^2$ is the metric on a unit two-sphere and $\mu$ is a mass
parameter. For $ 0 < \mu < \frac{1}{3} \Lambda^{-1/2} $, $ V $ has two
positive roots $r_{\rm c}$ and $r_{\rm b}$, corresponding to the
cosmological and the black hole horizons, respectively. The limit
where $\mu \rightarrow 0$ corresponds to the de~Sitter solution. In
the limit $\mu \rightarrow \frac{1}{3} \Lambda^{-1/2}$ the size of the
black hole horizon approaches the size of the cosmological horizon,
and the above coordinates become inappropriate, since $V(r)
\rightarrow 0$ between the two horizons. Following Ginsparg and
Perry~\cite{GinPer83}, we write
\begin{equation}
9 \mu^2 \Lambda = 1 - 3 \epsilon^2, \;\;
0 \leq \epsilon \ll 1. 
\end{equation}
Then the degenerate case corresponds to $ \epsilon \rightarrow 0 $. We
define new time and radial coordinates $ \psi $ and $ \chi $ by
\begin{equation}
\tau = \frac{1}{\epsilon\sqrt{\Lambda}} \psi; \;\;\;
r = \frac{1}{\sqrt{\Lambda}}
\left[1 - \epsilon\cos\chi - \frac{1}{6} \epsilon^2 \right].
\label{eq-transformations}
\end{equation}
In these coordinates the black hole horizon corresponds to $ \chi =
0 $ and the cosmological horizon to $ \chi = \pi $.  The new metric
obtained from the transformations is, to first order in $\epsilon$,
\begin{eqnarray}
ds^2 & = & - \frac{1}{\Lambda} \left( 1 +
                  \frac{2}{3}\epsilon\cos\chi
 \right) \sin^2\!\chi \; d\psi^2
     + \frac{1}{\Lambda} \left( 1 -
 \frac{2}{3}\epsilon\cos\chi \right)
                                 d\chi^2
\\ \nonumber
     & + & \frac{1}{\Lambda} \left( 1 -
 2\epsilon\cos\chi \right) d\Omega_2^2.
\label{eq-metric-eps}
\end{eqnarray}
This metric describes Schwarzschild-de~Sitter solutions of nearly
maximal black hole size.

In these coordinates the topology of the spacelike sections of
Schwarzschild-de~Sitter becomes manifest: $S^1 \times S^2$. In
general, the radius, $r$, of the two-spheres varies along the $S^1$
coordinate, $\chi$, with the minimal (maximal) two-sphere
corresponding to the black hole (cosmological) horizon. In the
degenerate case, the two-spheres all have the same radius.

\subsection{Thermodynamics}

The surface gravities of the two horizons are given by~\cite{BouHaw96}
\begin{equation}
  \kappa_{\rm c,\:b} = \sqrt{\Lambda}
    \left(1 \mp \frac{2}{3}\epsilon \right) + O( \epsilon^2),
  \label{eq-kappa-eps}
\end{equation}
where the upper (lower) sign is for the cosmological (black hole)
horizon. In the degenerate case, the two horizons have the same
surface gravity, and, since $T=\kappa/2\pi$, the same temperature.
They are in thermal equilibrium; one could say that the black hole
loses as much energy due to evaporation as it gains due to the
incoming radiation from the cosmological horizon. Away from the
thermal equilibrium, for nearly degenerate Schwarzschild-de~Sitter
black holes, one could make the simplifying assumption that the
horizons still radiate thermally, with temperatures proportional to
their surface gravities. This would lead one to expect an instability:
By Eq.~(\ref{eq-kappa-eps}), the black hole will be hotter than the
cosmological horizon, and will therefore suffer a net loss of
radiation energy. To investigate this suspected instability, a
two-dimensional model is constructed below, in which one-loop terms
are included.

\sect{Two-dimensional Model} \label{sec-2d-model}

The four-dimensional Lorentzian Einstein-Hilbert action with a
cosmological constant is
\begin{equation}
S = \frac{1}{16 \pi} \int d^4\!x\, (-g^{{\rm IV}})^{1/2} \left[
 R^{{\rm IV}} - 2 \Lambda - \frac{1}{2} \sum_{i=1}^{N}
 (\nabla^{{\rm IV}} f_i)^2 \right],
\end{equation}
where $R^{{\rm IV}}$ and $g^{{\rm IV}}$ are the four-dimensional
Ricci scalar and metric determinant, and the $f_i$ are scalar fields
which will carry the quantum radiation.

We shall consider only spherically symmetric fields and quantum
fluctuations. Thus, we make a spherically symmetric metric ansatz,
\begin{equation}
ds^2 = e^{2\rho} \left( -dt^2 + dx^2 \right) + e^{-2\phi} d\Omega^2,
\end{equation}
where the remaining two-dimensional metric has been written in
conformal gauge; $x$ is the coordinate on the one-sphere and has the
period $2\pi$. Now the spherical coordinates can be integrated out,
and the action is reduced to
\begin{equation}
S = \frac{1}{16\pi} \int d^2\!x\, (-g)^{1/2} e^{-2\phi} \left[
 R + 2 (\nabla \phi)^2 + 2 e^{2\phi} - 2
 \Lambda - \sum_{i=1}^{N} (\nabla f_i)^2 \right],
\end{equation}
where the gravitational coupling has been rescaled into the standard
form. Note that the scalar fields have acquired an exponential
coupling to the dilaton in the dimensional reduction. In order to take
quantum effects into account, we will find the classical solutions to
the action $S+W^*$. $W^*$ is the scale-dependent part of the one-loop
effective action for dilaton coupled scalars, which we derived in a
recent paper~\cite{BouHaw97a}:
\begin{equation}
W^* = - \frac{1}{48\pi} \int d^2\!x (-g)^{1/2} \left[ \frac{1}{2}
  R \frac{1}{\Box} R - 6 (\nabla \phi)^2 \frac{1}{\Box} R
  - 2 \phi R \right].
\end{equation}
The $(\nabla \phi)^2$ term will be neglected; we justify this neglect
at an appropriate place below.

Following Hayward~\cite{Hay94}, we render this action local by
introducing an independent scalar field $Z$ which mimics the trace
anomaly. The $f$ fields have the classical solution $f_i=0$ and can be
integrated out. Thus we obtain the action
\begin{eqnarray}
S\!\!\! &
 =\!\!\! & \frac{1}{16\pi} \int d^2\!x\, (-g)^{1/2} \left[ \left( 
e^{-2\phi} + \frac{ \kappa }{2} (Z +  w  \phi) \right) R
\right. \nonumber \\
& & \left. \mbox{\hspace{5em}}
 - \frac{ \kappa }{4} \left( \nabla Z \right)^2
+ 2 + 2 e^{-2\phi} \left( \nabla \phi \right)^2
- 2 e^{-2\phi} \Lambda \right]\!,
\end{eqnarray}
where
\begin{equation}
\kappa \equiv \frac{2N}{3}.
\end{equation}
There is some debate about the coefficient of the $\phi R$ term in the
effective action. Our result~\cite{BouHaw97a} corresponds to the
choice $ w =2$; the RST coefficient~\cite{RST1} corresponds to $ w
=1$, and the result of Nojiri and Odintsov~\cite{NojOdi97} can be
represented by choosing $ w =-6$.  In Ref.~\cite{Hay94}, probably
erroneously, $ w =-1$ was chosen.  We take the large $N$ limit, in
which the quantum fluctuations of the metric are dominated by the
quantum fluctuations of the $N$ scalars; thus, $ \kappa \gg 1$. In
addition, for quantum corrections to be small we assume that $ b
\equiv \kappa \Lambda \ll 1 $. To first order in $b$, we shall find
that the behaviour of the system is independent of $ w $.

For compactness of notation, we denote differentiation with respect to
$t$ ($x$) by an overdot (a prime). Further, we define for any
functions $f$ and $g$:
\begin{equation}
\partial f\,\partial g  \equiv - \dot{f} \dot{g} + f' g',\ \ \ \ 
\partial^2 g \equiv - \ddot{g} + g'',
\end{equation}
and
\begin{equation}
\delta f\,\delta g \equiv \dot{f} \dot{g} + f' g',\ \ \ \
\delta^2 g \equiv \ddot{g} + g''.
\end{equation}
Variation with respect to $\rho$, $\phi$ and $Z$ leads to the
following equations of motion:
\begin{equation}
- \left( 1 - \frac{  w \kappa }{4} e^{2\phi} \right)
 \partial^2 \phi + 2
(\partial \phi)^2 + \frac{ \kappa }{4} e^{2\phi} \partial^2 Z +
e^{2\rho+2\phi} \left( \Lambda e^{-2\phi} - 1 \right) = 0;
\label{eq-m-rho}
\end{equation}
\begin{equation}
\left( 1 - \frac{  w \kappa }{4} e^{2\phi}
 \right) \partial^2 \rho -
\partial^2 \phi + (\partial \phi)^2 + \Lambda e^{2\rho} = 0; 
\label{eq-m-phi}
\end{equation}
\begin{equation}
\partial^2 Z - 2 \partial^2 \rho = 0.
\label{eq-m-Z}
\end{equation}
There are two equations of constraint:
\begin{equation}
\left( 1 - \frac{  w \kappa }{4} e^{2\phi} \right)
 \left( \delta^2 \phi - 2 \delta\phi\,\delta\rho \right) -
(\delta\phi)^2
= \frac{\kappa}{8} e^{2\phi} \left[ (\delta Z)^2 + 2 \delta^2 Z
  - 4 \delta Z \delta\rho \right];
\label{eq-c1}
\end{equation}
\begin{equation}
\left( 1 - \frac{  w \kappa }{4} e^{2\phi} \right)
 \left( \dot{\phi}' - 
\dot{\rho} \phi' - \rho' \dot{\phi} \right) - \dot{\phi} \phi'
= \frac{\kappa}{8} e^{2\phi} \left[ \dot{Z} Z' + 2
\dot{Z}' - 2 \left( \dot{\rho} Z' + \rho'
  \dot{Z} \right) \right].
\label{eq-c2}
\end{equation}
From Eq.~(\ref{eq-m-Z}), it follows that
\begin{equation}
Z = 2\rho + \eta,
\label{eq-Z-eta}
\end{equation}
where $\eta$ satisfies
\begin{equation}
\partial^2 \eta = 0.
\label{eq-eta-laplace}
\end{equation}

The remaining freedom in $\eta$ can be used to satisfy the constraint
equations for any choice of $\rho$, $ \dot{\rho} $, $\phi$ and
$\dot{\phi}$ on an initial spacelike section. This can be seen most
easily by decomposing the fields and the constraint equations into
Fourier modes on the initial $S^1$. By solving for the second term on
the right hand side of Eq.~(\ref{eq-c1}), and by using
Eqs.~(\ref{eq-Z-eta}) and (\ref{eq-eta-laplace}), the first constraint
yields one algebraic equation for each Fourier coefficient of $\eta$.
Similarly, the second constraint yields one algebraic equation for the
time derivative of each Fourier coefficient of $\eta$.  If the initial
slice was non-compact, this argument would suffice.  Here it must be
verified, however, that $\eta$ and $\dot{\eta}$ will have a period of
$2\pi$.  The problem reduces to the question whether the two constant
mode constraint equations can be satisfied.  Indeed, while for each
oscillatory mode of $\eta$, there are two degrees of freedom (the
Fourier coefficient and its time derivative), the second time
derivative of the constant mode coefficient, $\ddot{\eta}_0$, must
vanish by Eq.~(\ref{eq-eta-laplace}). Thus there is only one degree of
freedom, $\dot{\eta}_0$, for the two constant mode equations.
However, since we have introduced no odd modes (i.e., modes of the
form $\sin kx$) in the perturbation of $\phi$, none of the fields will
contain any odd modes. Since each term in Eq.~(\ref{eq-c2}) contains
exactly one spatial derivative, each term will be odd.  Therefore all
even mode components of the second constraint vanish identically. In
particular the constant mode component will thus be automatically
satisfied. Then the freedom in $\dot{\eta}_0$ can be used to satisfy
the constant mode component of the remaining constraint,
Eq.~(\ref{eq-c1}), through the first\footnote{Note that $ \dot{\eta}_0
  $ can thus be purely imaginary, as indeed it will be for the Nariai
  solution, signaling negative energy density of the quantum field.}
term on the right hand side.

\sect{Perturbative Stability} \label{sec-stability}

\subsection{Perturbation Ansatz}

With the model developed above we can describe the quantum behaviour
of a cosmological black hole of the maximal mass under perturbations.
The Nariai solution is still characterised by the constancy of the
two-sphere radius, $e^{-\phi}$. Because of quantum corrections, this
radius will no longer be exactly $ \Lambda ^{-1/2}$.  Instead, the
solution is given by
\begin{equation}
e^{2\rho} = \frac{1}{\Lambda_1} \frac{1}{\cos^2\! t}, \ \ \ 
e^{2\phi} = \Lambda_2,
\label{eq-nariai}
\end{equation}
where
\begin{eqnarray}
\frac{1}{\Lambda_1} \!\!\! & = \!\!\! & \frac{1}{8 \Lambda} \left[
  4 - (w\!+\!2)b + \sqrt{16 - 8(w\!-\!2)b + (w\!+\!2)^2 b^2}
 \right]\!; \\
\Lambda_2 \!\!\! & = \!\!\! & \frac{1}{2 w\kappa} \left[ 
  4 + (w\!+\!2)b - \sqrt{16 - 8(w\!-\!2)b + (w\!+\!2)^2 b^2}
 \right]\!.
\end{eqnarray}
Expanding to first order in $b$, one obtains:
\begin{eqnarray}
\frac{1}{\Lambda_1} & \approx &
  \frac{1}{\Lambda} \left( 1 - \frac{wb}{4} \right); \\
\Lambda_2 & \approx &  \Lambda \left( 1 - \frac{b}{2} \right).
\end{eqnarray}

Let us now perturb this solution so that the two-sphere radius,
$e^{-\phi}$, varies slightly along the one-sphere coordinate, $ x$:
\begin{equation}
e^{2\phi} =  \Lambda_2 \left[ 1 + 2 \epsilon
 \sigma(t) \cos x \right],
\label{eq-pert-phi}
\end{equation}
where we take $ \epsilon \ll 1$. We will call $\sigma$ the {\em metric
  perturbation}. A similar perturbation could be introduced for
$e^{2\rho}$, but it does not enter the equation of motion for $\sigma$
at first order in $ \epsilon $. This equation is obtained by
eliminating $\partial^2 Z$ and $\partial^2 \rho$ from
Eq.~(\ref{eq-m-rho}) using Eqs.~(\ref{eq-m-Z}) and~(\ref{eq-m-phi}),
and inserting the above perturbation ansatz. This yields
\begin{equation}
\frac{\ddot{\sigma}}{\sigma} = \frac{a}{\cos^2\! t} - 1,
\label{eq-m-si}
\end{equation}
where
\begin{equation}
a \equiv \frac{2 \sqrt{16 - 8(w\!-\!2)b + (w\!+\!2)^2 b^2}}{4-wb}
\end{equation}
To first order in $b$, one finds that
\begin{equation}
a \approx 2+b,
\end{equation}
which means that $w$, and therefore the $\phi R$ term in the effective
action, play no role in the horizon dynamics at this level of
approximation. This is also the right place to discuss why the term
$\sqrt{-g}\, (\nabla \phi)^2 \frac{1}{\Box} R$ in the effective action
can be neglected. In conformal coordinates this term is proportional
to $(\partial \phi)^2 \rho$. Thus, in the $\rho$-equation of motion,
Eq.~(\ref{eq-m-rho}), it will lead to a $(\partial \phi)^2$ term,
which is of second order in $ \epsilon $ and can be neglected.  In the
$\phi$-equation of motion, Eq.~(\ref{eq-m-phi}), it yields terms
proportional to $\kappa$ that are of first order in $ \epsilon $. They
will enter the equation of motion for $\sigma$ via the $\kappa
e^{2\phi} \partial^2 Z$ term in Eq.~(\ref{eq-m-phi}). Thus they will
be of second order in $b$ and can be dropped. The neglect of the $\log
\mu^2$ term~\cite{BouHaw97a} can be justified in the same way.

\subsection{Horizon Tracing}

In order to describe the evolution of the black hole, one must know
where the horizon is located. The condition for a horizon is
$(\nabla \phi)^2 = 0$. Eq.~(\ref{eq-pert-phi}) yields
\begin{equation}
\frac{\partial\phi}{\partial t} = \epsilon \dot{\sigma} \cos x,\ \ \
\frac{\partial\phi}{\partial x} = - \epsilon \sigma \sin x.
\end{equation}
Therefore, the black hole and cosmological horizons are located at
\begin{equation}
 x_{{\rm b}}(t) = \arctan \left| \frac{\dot{\sigma}}{\sigma}
 \right|, \ \ \  x_{{\rm c}}(t) = \pi -  x_{{\rm b}}(t).
\label{eq-chib}
\end{equation}
To first order in $ \epsilon $, the size of the black hole horizon,
$r_{{\rm b}}$, is given by
\begin{equation}
r_{{\rm b}}(t)^{-2} = e^{2\phi[t, x_{{\rm b}}(t)]} = 
\Lambda_2 \left[ 1 + 2 \epsilon \delta(t) \right],
\label{eq-rb}
\end{equation}
where we define the {\em horizon perturbation}
\begin{equation}
\delta \equiv \cos  x_{\rm b} = \sigma \left( 1+
 \frac{\dot{\sigma}^2}{\sigma^2} \right)^{-1/2}.
\label{eq-delta}
\end{equation}
We will focus on the early time evolution of the black hole horizon;
the treatment of the cosmological horizon is completely equivalent.

To obtain explicitly the evolution of the black hole horizon radius,
$r_{{\rm b}}(t)$, one must solve Eq.~(\ref{eq-m-si}) for $ \sigma(t)$,
and use the result in Eq.~(\ref{eq-delta}) to evaluate
Eq.~(\ref{eq-rb}). If the horizon perturbation grows, the black hole
is shrinking: this corresponds to evaporation. It will be shown below,
however, that the behaviour of $\delta(t)$ depends on the initial
conditions chosen for the metric perturbation, $ \sigma_0$ and $
\dot{\sigma}_0$.

\subsection{Classical Evolution}

As a first check, one can examine the classical case, $ \kappa=0$.
This has $a=2$, and Eq.~(\ref{eq-m-si}) can be solved exactly. From
the constraint equations, Eq.~(\ref{eq-c1}) and (\ref{eq-c2}), it
follows that
\begin{equation}
\dot{\sigma} =\sigma\, \tan t.
\end{equation}
Therefore the appropriate boundary condition at $t=0$ is $
\dot{\sigma}_0=0$. The solution is
\begin{equation}
\sigma(t) = \frac{\sigma_0}{\cos t}.
\end{equation}
Then Eq.~(\ref{eq-delta}) yields
\begin{equation}
\delta(t) = \sigma_0 = {{\rm const}}.
\end{equation}
Since the quantum fields are switched off, no evaporation takes place;
the horizon size remains that of the initial perturbation. This simply
describes the case of a static Schwarzschild-de~Sitter solution of
nearly maximal mass, as given in Eq.~(\ref{eq-metric-eps}).

\subsection{Quantum Evolution}

When we turn on the quantum radiation ($ \kappa > 0$) the constraints
no longer fix the initial conditions on the metric perturbation. There
will thus be two linearly independent types of initial perturbation.
The first is the one we were forced to choose in the classical case: $
\sigma_0 > 0$, $ \dot{\sigma}_0=0$.  It describes the spatial section
of a quantum corrected Schwarzschild-de~Sitter solution of nearly
maximal mass.  Thus one might expect the black hole to evaporate. For
$a>2$, Eq.~(\ref{eq-m-si}) cannot be solved analytically. Since we are
interested in the early stages of the evaporation process, however, it
will suffice to solve for $ \sigma$ as a power series in $t$. Using
Eq.~(\ref{eq-delta}) one finds that
\begin{eqnarray}
\delta(t) & = & \sigma_0 \left[ 1 - \frac{1}{2}(a-1)(a-2) t^2 + O(t^4)
\right] \nonumber \\
& \approx & \sigma_0 \left[ 1 - \frac{1}{2}bt^2 \right].
\end{eqnarray}
The horizon perturbation shrinks from its initial value.  Thus, the
black hole size {\em increases}, and the black hole grows, at least
initially, back towards the maximal radius. One could say that nearly
maximal Schwarzschild-de~Sitter black holes ``anti-evaporate''.

This is a surprising result, since intuitive thermodynamic arguments
would have led to the opposite conclusion. The anti-evaporation can be
understood in the following way. By specifying the metric
perturbation, the radiation distribution of the $Z$ field is
implicitly fixed through the constraint equations, (\ref{eq-c1})
and~(\ref{eq-c2}). Our result shows that radiation is heading towards
the black hole if the boundary condition $ \sigma_0 > 0$, $
\dot{\sigma}_0=0$ is chosen.

Let us now turn to the second type of initial metric perturbation: $
\sigma_0=0$, $ \dot{\sigma}_0 > 0$. Here the spatial geometry is
unperturbed on the initial slice, but it is given a kind of ``push''
that corresponds to a perturbation in the radiation bath. Solving once
again for $ \sigma $ with these boundary conditions, and using
Eq.~(\ref{eq-delta}), one finds for small $t$:
\begin{equation}
\delta(t) = \dot{\sigma}_0 t^2.
\end{equation}
The horizon perturbation grows. This perturbation mode is unstable,
and leads to evaporation.

We have seen that the radiation equilibrium of a Nariai universe
displays unusual and non-trivial stability properties. The evolution
of the black hole horizon depends crucially on the type of metric
perturbation. Nevertheless, one may ask the question whether a
cosmological black hole will typically evaporate or not. Cosmological
black holes cannot come into existence through classical gravitational
collapse, since they live in an exponentially expanding de~Sitter
background. The only natural way for them to appear is through the
quantum process of pair creation~\cite{GinPer83}. This pair creation
process can also occur in an inflationary universe, because of its
similarity to de~Sitter space~\cite{BouHaw96,BouHaw95,Bou96}. The
nucleation of a Lorentzian black hole spacetime is described as the
analytic continuation of an appropriate complex solution of the
Einstein equations, which satisfies the no boundary
condition~\cite{HarHaw83}. We will show below that the no boundary
condition selects a particular linear combination of the two types of
initial metric perturbation, thus allowing us to determine the fate of
the black hole.

\sect{No Boundary Condition} \label{sec-nbp}

To obtain the unperturbed Euclidean Nariai solution in conformal
gauge, one performs the analytic continuation $t = i\tau$ in the
Lorentzian solution, Eq.~(\ref{eq-nariai}). This yields
\begin{equation}
\left( ds^{{\rm IV}} \right)^2 = e^{2\rho} \left( d\tau^2 +
  d x^2 \right) + e^{-2\phi} d\Omega^2,
\end{equation}
and
\begin{equation}
e^{2\rho} = \frac{1}{\Lambda_1} \frac{1}{\cosh^2\! \tau},\ \ \
e^{2\phi} = \Lambda_2.
\end{equation}
In four dimensions, this describes the product of two round
two-spheres of slightly different radii, $ \Lambda_1^{-1/2}$ and $
\Lambda_2^{-1/2}$. The analytic continuation to a Lorentzian Nariai
solution corresponds to a path in the $\tau$ plane, first along the
real $\tau$ axis, from $\tau = -\infty$ to $\tau=0$, and then along
the imaginary axis from $t=0$ to $t=\pi/2$. This can be visualised
geometrically by cutting the first two sphere in half, and joining to
it a Lorentzian $1+1$-dimensional de~Sitter hyperboloid.  Because the
$(\tau, x)$ sphere has its north (south) pole at $\tau=\infty$
($\tau=-\infty$), it is convenient to rescale time:
\begin{equation}
\sin u = \frac{1}{\cosh \tau},
\end{equation}
or, equivalently,
\begin{equation}
\cos u = - \tanh \tau,\ \ \ \cot u = -\sinh \tau, \ \ \ 
du = \frac{d\tau}{\cosh \tau}.
\end{equation}
With the new time coordinate $u$, the solution takes the form
\begin{equation}
\left( ds^{{\rm IV}} \right)^2 =  \frac{1}{\Lambda_1} \left( du^2 +
  \sin^2\! u\, d x^2 \right) +  \frac{1}{\Lambda_2} d\Omega^2.
\end{equation}
Now the south pole lies at $u=0$, and the nucleation path runs to
$u=\pi/2$, and then parallel to the imaginary axis
($u=\pi/2+iv$) from $v=0$ to $v=\infty$.

The perturbation of $e^{2\phi}$, Eq.~(\ref{eq-pert-phi}) introduces
the variable $\sigma$, which satisfies the Euclidean version of
Eq.~(\ref{eq-m-si}):
\begin{equation}
\sin^2\! u\,
 \frac{d^2 \sigma}{du^2} + \sin u \cos u \frac{d \sigma}{du} -
\left( 1-a \sin^2\! u\, \right) \sigma = 0.
\label{eq-m-si-eucl}
\end{equation}
In addition, the nature of the Euclidean geometry enforces the
boundary condition that the perturbation vanish at the south pole:
\begin{equation}
\sigma(u=0) = 0.
\label{eq-nbp-si}
\end{equation}
Otherwise, $e^{2\phi}$ would not be single valued, because the
coordinate $ x$ degenerates at this point. This leaves $ \dot{\sigma}
$ as the only degree of freedom in the boundary conditions at $u=0$.

It will be useful to define the parameter $c$ by the relation $c(c+1)
\equiv a$. The classical case, $a=2$, corresponds to $c=1$; for small
$b$, they receive the quantum corrections $a=2+b$ and $c=1+b/3$. With
the boundary condition, Eq.~(\ref{eq-nbp-si}), the equation of motion
for $ \sigma $, Eq.~(\ref{eq-m-si-eucl}), can be solved exactly only
for integer $c$ ($a=2$, 6, 12, 20,\ldots). The solution is of the form
\begin{equation}
\sigma( u) = \sum_{0 \leq k < c/2} A_k\, \sin(c-2k) u,
\label{eq-sigma-u}
\end{equation}
with constants $A_k$. Even for non-integer $c$, however, this turns
out to be a good approximation in the region $0 \leq u \leq \pi/2$ of
the $(u,v)$ plane. Since we are interested in the case where $b \ll
1$, the sum in Eq.~(\ref{eq-sigma-u}) contains only one term, and we
use the approximation\footnote{Treating Eq.~(\ref{eq-m-si-eucl})
  perturbatively in $b$ around $a=2$ leads to untractable integrals.
  Fortunately the guessed approximation in Eq.~(\ref{eq-si-eucl})
  turns out to be rather accurate, especially for late Lorentzian
  times $v$, which is the regime in which we claim our results to be
  valid. It is easy to check numerically that for sufficiently large
  $v$ ($v>10$), both the real and the imaginary part of
  Eq.~(\ref{eq-si-eucl}) have a relative error $b/30$ or less. The
  result for the phase of the prefactor, Eq.~(\ref{eq-prefactor}), has
  a relative error of less than $10^{-4}$, independently of $b$.
  Crucially, the exponential behaviour at late Lorentzian times is
  reproduced perfectly, as the ratio \[ \frac{\partial \sigma/\partial
    v}{\sigma}, \] using the approximation, agrees with the numerical
  result to machine accuracy.  Therefore the relative error in
  Eq.~(\ref{eq-delta-result}) is the same as in
  Eq.~(\ref{eq-si-eucl}); in both equations it is located practically
  entirely in the magnitude of the prefactor. --- These statements
  hold for $0\leq b \leq 1$, which really is a wider interval than
  necessary.}
\begin{equation}
\sigma( u) \approx \tilde{A} \sin c u.
\label{eq-si-eucl}
\end{equation}

It is instructive to consider the classical case first. (Physically,
this is questionable, since the no boundary condition violates the
constraints at second order in $ \epsilon $.) For $b=0$, the solution
$\sigma( u) = \tilde{A} \sin u$ is exact. Along the Lorentzian line
($u=\pi/2+iv$), this solution becomes $\sigma(v) = \tilde{A} \cosh v$.
By transforming back to the Lorentzian time variable $t$, one can
check that this is the stable solution found in the previous section,
with $ \sigma_0 = \tilde{A} $, $ \dot{\sigma}_0=0$. For real
$\tilde{A}$, it is real everywhere along the nucleation path. Thus,
when the quantum fields are turned off, the Euclidean formalism
predicts that the unstable mode will not be excited. This is a welcome
result, since there are no fields that could transport energy from one
horizon to another.

Once $b$ is non-zero, however, it is easy to see that $\partial
\sigma/\partial u$ no longer vanishes at the origin of Lorentzian
time, $u=\pi/2$. This indicates that the unstable mode, $
\dot{\sigma}_0 \neq 0$, will be excited. Unfortunately, checking this
is not entirely straightforward, because $\sigma$ is no longer real
everywhere along the nucleation path. One must impose the condition
that $\sigma$ and $ \dot{\sigma} $ be real at late Lorentzian times.
We will first show that this can be achieved by a suitable complex
choice of $A$. One can then calculate the horizon perturbation,
$\delta$, from the real late-time evolution of the metric
perturbation, $ \sigma $, to demonstrate that evaporation takes place.

From Eq.~(\ref{eq-si-eucl}) one obtains the Lorentzian evolution of
$\sigma$,
\begin{eqnarray}
\sigma(v) & = & \tilde{A} \sin c \left( \frac{\pi}{2} + iv \right) \\
& = & \tilde{A} \left( \sin \frac{c\pi}{2} \cosh cv
  + i \cos \frac{c\pi}{2} \sinh cv \right).
\end{eqnarray}
For late Lorentzian times (i.e., large $v$), $\cosh cv \approx \sinh
cv \approx e^{cv}/2$, so the equation becomes
\begin{equation}
\sigma(v) \approx \frac{1}{2} \tilde{A} \left( i e^{-ic\pi/2}
                  \right) e^{cv}.
\end{equation}
This can be rendered purely real by choosing the complex constant
$\tilde{A}$ to be
\begin{equation}
\tilde{A} = A \left( -i e^{ic\pi/2} \right),
\label{eq-prefactor}
\end{equation}
where $A$ is real.

Now we can return to the question whether the Euclidean boundary
condition leads to evaporation. After transforming the time
coordinate, the expression for the growth of the horizon perturbation,
Eq.~(\ref{eq-delta}), becomes
\begin{equation}
\delta(v) = \sigma \left[ 1+ \cosh^2\! v \left(
 \frac{\partial \sigma/\partial v}{\sigma} \right)^2 \right]^{-1/2}.
\label{eq-delta-v}
\end{equation}
The late time evolution is given by $ \sigma(v) = \frac{A}{2}
e^{cv} $. This yields, for large $v$,
\begin{equation}
\delta(v) \approx \frac{A}{2}
e^{cv} \left( 1+c^2 e^{2v} \right)^{-1/2} \approx
\frac{A}{2c} \exp(\frac{b}{3} v).
\label{eq-delta-result}
\end{equation}
This result confirms that pair created cosmological black holes will
indeed evaporate. The magnitude of the horizon perturbation is
proportional to the initial perturbation strength, $A$. The
evaporation rate grows with $\kappa\Lambda$. This agrees with
intuitive expectations, since $\kappa$ measures the number of quantum
fields that carry the radiation.

\sect{Summary}

We have investigated the quantum stability of the
Schwarzschild-de~Sitter black holes of maximal mass, the Nariai
solutions. From four-dimensional spherically symmetric general
relativity with a cosmological constant and $N$ minimally coupled
scalar fields we obtained a two-dimensional model in which the scalars
couple to the dilaton.  The one-loop terms were included in the large
$N$ limit, and the action was made local by introducing a field $Z$
which mimics the trace anomaly.

We found the quantum corrected Nariai solution and analysed its
behaviour under perturbations away from degeneracy. There are two
possible ways of specifying the initial conditions for a perturbation
on a Lorentzian spacelike section. The first possibility is that the
displacement away from the Nariai solution is non-zero, but its time
derivative vanishes. This perturbation corresponds to nearly
degenerate Schwarzschild-de~Sitter space, and somewhat surprisingly,
this perturbation is stable at least initially. The second possibility
is a vanishing displacement and non-vanishing derivative. These
initial conditions lead directly to evaporation. The different
behaviour of these two types of perturbations can be explained by the
fact that the initial radiation distribution is implicitly specified
by the initial conditions, through the constraint equations.

If neutral black holes nucleate spontaneously in pairs on a de~Sitter
background, the initial data will be constrained by the no boundary
condition: it selects a linear combination of the two types of
perturbations. By finding appropriate complex compact instanton
solutions we showed that this condition leads to black hole
evaporation. Thus neutral primordial black holes are unstable.

\end{document}